\documentclass[dvips]{article}
\usepackage{icrctc07}

\newcommand{\beq}{\begin{equation}}
\newcommand{\eeq}{\end{equation}}

\newcommand{\be}{\begin{equation}}
\newcommand{\ee}{\end{equation}}

\title{Large-Angular-Scale Clustering as a Clue to the Source of UHECRs}

\shorttitle{Clustering of UHECRs}

\authors{Andreas A. Berlind, Glennys R. Farrar}
\shortauthors{Berlind & Farrar}

\afiliations{Center For Cosmology and Particle Physics, 
Department of Physics\\ 
New York University, New York, NY 10003, USA}
\email{aberlind@cosmo.nyu.edu,farrar@physics.nyu.edu}

\abstract{We show that future Ultra-High Energy Cosmic Ray samples should be able to 
distinguish whether the sources of UHECRs are hosted by galaxy clusters or ordinary 
galaxies, or whether the sources are uncorrelated with the large-scale structure of 
the universe.  Moreover, this is true independently of arrival direction uncertainty 
due to magnetic deflection or measurement error.  The reason for this is the simple 
property that the strength of large-scale clustering for extragalactic sources depends 
on their mass, with more massive objects, such as galaxy clusters, clustering more 
strongly than lower mass objects, such as ordinary galaxies.
}

\begin{document}
\maketitle

\section{Introduction} \label{sec:intro}
Identifying the sources of ultrahigh energy cosmic rays (UHECRs, here $E>10^{19}$eV
$\equiv 10$ EeV) is complicated by the deflection they presumably experience in Galactic
and extragalactic magnetic fields, as well as their relatively poor arrival direction
determinations, typically $\sim 1^{\circ}$.  Arrival directions of most UHECRs are thus
not known well enough to match their positions with specific astrophysical objects.
However, there is also useful information in the clustering of UHECRs on large scales,
where $\sim$ few degree uncertainties in position become unimportant.  The clustering of
galaxies in the universe is typically quantified by the two-point correlation function
or its analog in Fourier space, the power spectrum.  The two-point correlation function
$\xi(r)$ of any class of objects (e.g., galaxies of a certain luminosity or color) is
defined as the excess number of pairs of such objects at physical separation $r$ over
that expected for a random (Poisson) distribution. In Cold Dark Matter models, the
large-scale amplitude of $\xi(r)$ (usually referred to as the bias) of a population of
objects depends only on their mass, with more massive objects, such as clusters of
galaxies, clustering more strongly than less massive objects, such as ordinary galaxies
\cite{mo_white_96,sheth_tormen_99,berlind_etal_06b}.
The large-scale bias of a UHECR sample is therefore a robust and informative measure of
the clustering properties of the source.  We cannot measure physical separations for
pairs involving UHECRs because they do not have measured redshifts.  However, we can
measure the angular correlation function $\omega(\theta)$.  As is the case for $\xi(r)$,
the large-scale amplitude of $\omega(\theta)$ for a UHECR sample depends on the nature of
the astrophysical source.  However, it also depends on the depth of the sample because
deeper samples mix more physically uncorrelated pairs and thus show weaker angular
clustering.  In order to access the information in the large-scale angular clustering of
UHECRs, we must therefore know the depth of our UHECR sample.  In this paper, we
demonstrate what can be learned from the large-scale angular clustering of UHECRs, we
estimate what kind of sample is needed to do this analysis, and we show how to deal with
the unknown depth of a UHECR sample, using the GZK effect.

\section{Large-Angle Clustering of UHECRs} \label{sec:biasCR}

We demonstrate what can be learned from the large-angle clustering of UHECRs by creating
mock samples of UHECRs assuming different astrophysical sources and examining their
resulting clustering.  We use the Sloan Digital Sky Survey (SDSS) \cite{york_etal_00}
to create a volume-limited sample of galaxies that is complete out to a distance of 
286Mpc.  We select a sample of massive galaxy clusters in the same volume taken from a
SDSS group and cluster catalog \cite{berlind_etal_06a}.  Based on their luminosities,
we estimate these clusters to have masses greater than $10^{14} h^{-1}M_{\odot}$.
We then measure angular cross-correlation functions of each of these samples with the 
galaxy sample (so, for the galaxy case, we are measuring the autocorrelation) using the
Landy-Szalay \cite{landy_szalay_93} estimator:
$$
\omega_{12}(\theta) = \frac{N_{D_1D_2} - N_{D_1R} - N_{D_2R} + N_{RR}}{N_{RR}},
$$
where $N_{D_1D_2}$ is the number of pairs as a function of $\theta$ between the two 
data samples (in this case, galaxies and something else), $N_{D_1R}$ and $N_{D_2R}$ are 
the number of pairs as a function of $\theta$ between each data sample and a random 
sample, and $N_{RR}$ is the number of random-random pairs.  Figure~1 shows the resulting 
angular correlation functions: cluster-galaxy, galaxy-galaxy, as well as the 
random-galaxy case.  As expected, the cluster-galaxy correlation function has a higher 
amplitude than the galaxy-galaxy correlation function on all angular scales, and the 
random-galaxy correlation function is equal to zero by construction.

\begin{figure}[t]
\begin{center}
\noindent
\includegraphics [width=0.5\textwidth]{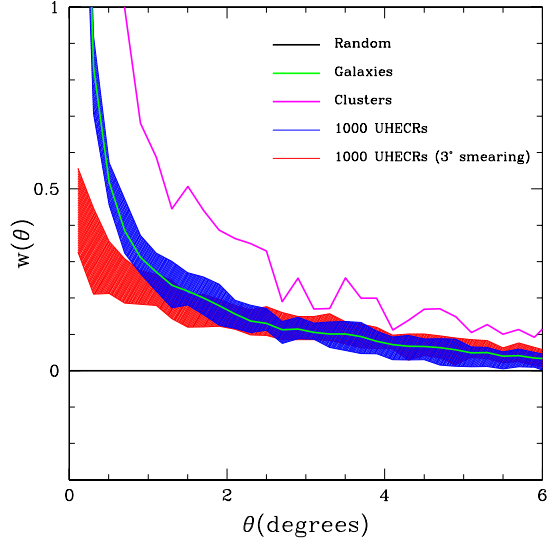}
\end{center}
\caption{
Predicted UHECR-galaxy angular cross-correlation functions for the cases that 
the astrophysical sources of UHECRs are (1) uncorrelated with the large-scale structure 
in the universe ({\it black line}), (2) ordinary galaxies ({\it green curve}), and (3) 
clusters of galaxies ({\it magenta curve}).  The blue shaded region shows the 95\% 
($2\sigma$) measurements using 200 mock samples of 1000 UHECRs each, where the UHECRs 
are assumed to originate in galaxies.  The red shaded region shows the same, but for 
mock UHECR arrival directions containing $3^\circ$ random Gaussian errors.  For this 
calculation, a SDSS galaxy sample of median depth 230Mpc was used.  
}\label{fig:1}
\end{figure}

These three curves represent predictions for the UHECR-galaxy cross-correlation function
in the three distinct cases that UHECRs originate from astrophysical sources that: (1)
live in massive clusters, (2) live in ordinary galaxies, and (3) are uncorrelated with
the large-scale structure of the universe, such as sources within the Milky Way galaxy.
The three cases predict different measured UHECR-galaxy correlation functions even at
large angles, where UHECR direction uncertainties due to measurement error and magnetic
deflections are unimportant.

We next examine how well we can distinguish between these different predictions assuming
a sample of 1000 UHECRs.  For the purpose of this test, we assume that the sources of
UHECRs are, in fact, ordinary galaxies. We create a mock UHECR sample by randomly
selecting 1000 galaxies from our SDSS galaxy sample.  We create 200 independent mock
samples in this way and measure their cross correlation with all galaxies.  The shaded
blue region in Figure~1 contains 95\% ($2\sigma$) of the mock realizations.  We then
simulate arrival direction uncertainties by applying a random $3^\circ$ Gaussian smearing
to all our mock UHECRs and repeating the correlation function measurements.  The red
shaded region in Figure~1 shows the 95\% dispersion for these new measurements.  As
expected, the $3^\circ$ smearing drastically reduces the correlation function at small
angular scales, but has a negligible effect on scales larger than $\sim 2^\circ$.
Figure~1 shows that with a sample of 1000 UHECRs, the measured clustering at large angles
($\geq4^\circ$) alone can easily distinguish between the ``cluster'', ``galaxy'', and 
``random'' hypotheses.

\section{What Kind of UHECR Sample Do We Need?} \label{sec:depth}

Although the sample of 1000 UHECRs used in Figure~1 is large compared to current
available samples, the sample depth in the above illustration is also large (median
depth=230Mpc).  The angular clustering will have a higher signal in shallower samples
because each angular bin will mix in fewer uncorrelated pairs, so we can get away with
smaller UHECR samples in shallower volumes.  We explore this in Figure~2, where we show
the signal-to-noise (S/N) of a measured UHECR-galaxy cross-correlation on large angular
scales ($6-8^\circ$), as a function of sample size $N_{\mathrm{CR}}$ and depth.  In 
order to calculate this, we do the same sort of mock UHECR analysis as in Figure~1, but 
using galaxies from the 2MASS survey \cite{kleinmann_etal_94}.

Figure~2 shows that if we want a S/N=3 (99.7\% significance) detection of UHECRs
clustering like ordinary 2MASS galaxies, we need 40 UHECRs of median source-distance
$d_{\mathrm{med}}=50$Mpc, or $N_{\mathrm{CR}}=80$ with $d_{\mathrm{med}}=80$Mpc, or 
$N_{\mathrm{CR}}=160$ with $d_{\mathrm{med}}=110$Mpc, or $N_{\mathrm{CR}}=320$ with 
$d_{\mathrm{med}}=150$Mpc.

We now return to the issue of the unknown depth of a given UHECR sample.  Fortunately,
the GZK energy loss phenomenon provides a way to put a limit on the depth of a UHECR
sample. The rapid variation with energy of the energy loss means that an ensemble of
UHECRs of a given energy has a rather well-defined horizon within which they are
produced.  If we assume that the energies of UHECRs are well determined, we can use the
GZK effect to solve for the distance distribution of a UHECR sample, given an initial
energy spectrum of cosmic rays.  Assuming an $E^{-2.7}$ energy spectrum, we compute the 
median depth of an UHECR sample as a function of its lower energy cutoff, and show the 
result in Figure~3.  We can now use Figure~3 to connect the sample depths shown in 
Figure~2 with energy cutoffs for UHECR samples.  In our S/N=3 example, the required 
samples would have 40, 80, 160, and 320 UHECRs with energies above 90EeV, 56EeV, 45EeV, 
and 37EeV, respectively.  These samples are larger than currently available samples from
AGASA+HiRes, but should be available in the near future by the Pierre Auger experiment.

\begin{figure}[t]
\begin{center}
\noindent
\includegraphics [width=0.5\textwidth]{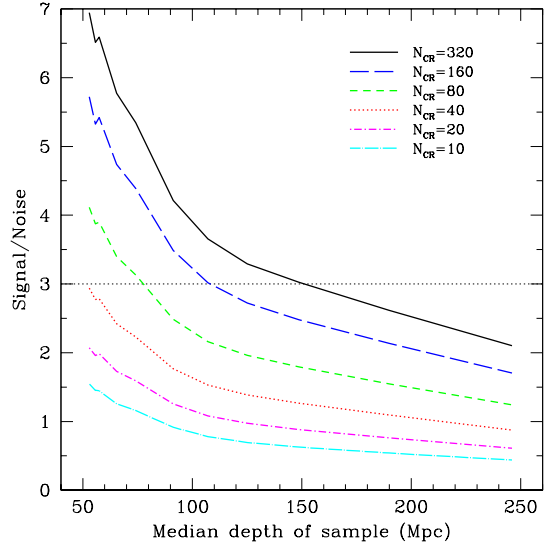}
\end{center}
\caption{
Estimated signal-to-noise (S/N) for measurements of the UHECR-galaxy
cross-correlation function on large angular scales ($6-8^\circ$) as a function of median
sample depth and size of UHECR sample.  This calculation was done using 2MASS galaxy
samples of various sample depths, and assuming that UHECRs originate from these same
galaxies.  Different colored curves represent different size UHECR samples, as listed in
the panel.  This plot answers the question:  At what significance can we detect the
cross-correlation between UHECRs and 2MASS galaxies at large angular scales, if we have
a UHECR sample of size $N_{\mathrm{CR}}$ and a given galaxy and UHECR sample depth?
}\label{fig:2}
\end{figure}

\begin{figure}[t]
\begin{center}
\noindent
\includegraphics [width=0.5\textwidth]{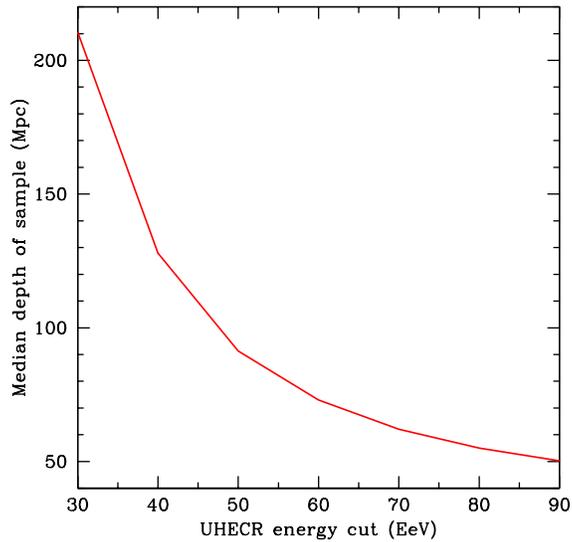}
\end{center}
\caption{
Median depth of a UHECR sample as a function of its lower energy cutoff, assuming
that the probability distribution of distances for a single UHECR is given by the GZK
effect.  This calculation was done assuming a UHECR energy spectrum of $E^{-2.7}$.  For
each energy threshold, a total distance distribution was computed by weighting the
probability distributions of individual energies by the overall energy spectrum.  The
sample depth decreases with energy because of the GZK effect.
}\label{fig:3}
\end{figure}

\section{Acknowledgements}

Funding for the creation and distribution of the SDSS has been provided by the  Alfred P. Sloan Foundation, the Participating Institutions, NASA, the NSF, the U.S. Department of Energy, the Japanese Monbukagakusho, and the Max Planck Society. The research of G. R. Farrar has been supported in part by NSF-PHY-0401232 and that of A. A. Berlind by the James Arthur Endowment of New York University, NSF-PHY-0401232 and NASA NAG5-9246.

\bibliography{icrc1259}
\bibliographystyle{plain}
\end{document}